\documentclass[doublecol]{epl2} 

\title{Phenomenological theory of the Potts model evaporation-condensation transition}
\shorttitle{} 

\author{M. Ib\'a\~nez-Berganza}
\shortauthor{M. Ib\'a\~nez-Berganza}

\institute{
 INFN-Gruppo Collegato di Parma, via G.P. Usberti, 7/A, 43124, Parma, Italy \\
 Dipartimento di Fisica, Universit\`a di Roma ``La Sapienza'', Piazzale Aldo Moro 5, 00185, Roma, Italy
}

\def\beq{\begin{equation}}
\def\eeq{\end{equation}}
\def\barray{\begin{eqnarray}}
\def\earray{\end{eqnarray}}
\def\<{\langle}
\def\>{\rangle}

\newcommand{\order}{{\cal O}}

\usepackage{amsmath,amssymb}
\def\e{\epsilon}
\def\d{{\rm d}}
\def\ed{\epsilon_{\rm d}}
\def\eo{\epsilon_{\rm o}}
\def\eb{\epsilon_{\rm b}}
\def\sd{s_{\rm d}}
\def\so{s_{\rm o}}
\def\sc{s_{\rm c}}
\def\betad{{\beta_{\rm d}}}
\def\cd{c_{\rm d}}
\def\s{\sigma}

\abstract{
We present a phenomenological theory describing the finite-size evaporation-condensation transition of the $q$-state Potts model in the microcanonical ensemble. Our arguments rely on the existence of an exponent $\s$, relating the surface and the volume of the condensed phase droplet. The evaporation-condensation transition temperature and energy converge to their infinite-size values with the same power, $a=(1-\s)/(2-\s)$, of the inverse of the system size. 
For the 2D Potts model we show, by means of efficient simulations up to $q=24$ and $1024^2$ sites, that the exponent $a$ is compatible with $1/4$, in disagreement with previous studies. While this value cannot be addressed by the evaporation-condensation theory developed for the Ising model, it is obtained in the present scheme if $\s=2/3$, in agreement with previous theoretical guesses. The connection with the  phenomenon of metastability in the canonical ensemble is also discussed.
}

\begin{document}

\maketitle

\section{Introduction}

The equilibrium and dynamics of phase coexistence in first-order phase transitions is a crucial and longstanding topic in statistical physics \cite{Binder1987Theory,Gunton1983Phase,Fisher1967Theory}. Since the eighties there has been a considerable interest in the so called micro-ensemble, in which the extensive thermodynamic variable, the magnetization in the para/ferromagnetic transition, or the particle density in the vapor/liquid case, is fixed. While systems with short-range interactions present ensemble equivalence in the thermodynamic limit, at finite sizes they exhibit a variety of {\it equilibrium} phenomenology in the micromagnetic or microcanonical ensembles, absent in the magnetic or canonical ensembles, as negative specific heat and reentrance of the entropy-temperature curve. 

Of remarkable relevance is the evaporation-condensation (EC) transition \cite{Binder1980Critical,Biskup2002,Neuhaus2003,Nussbaumer2006,Binder2003Theory,Nussbaumer2010}, that we now illustrate in the microcanonical ensemble.  Consider a system presenting a first-order transition at fixed energy density $\e$ in phase coexistence $\eo<\e<\ed$ ($\eo$, $\ed$ being the low and high  energies at the transition in the thermodynamic limit), whose thermodynamic potential is the entropy density $s$, satisfying $\d s=\beta \d\e$ (see fig. \ref{fig:quantities}). The situation is equivalent for magnets with the free energy, the magnetization $m$ and the magnetic field $h$, or for liquids with the free energy, the intensive volume, minus the pressure,  playing the role of $s$, $\e$ and $\beta$, respectively. In the thermodynamic limit the phase coexistence is given by the {\it lever rule}: for $\e \lesssim \ed$, the minority (low-$s$) phase  condensates in a connected {\it droplet} whose volume equals a fraction $(\ed-\e)/(\ed-\eo)$ of the system volume. In the presence of the droplet, the (high-$s$ phase) bulk energy surrounding it is larger than the global energy $\e$, hence its formation increases the entropy. This effect, however, is attenuated for finite sizes for which the droplet/bulk interface energy is not negligible and lowers the bulk energy.  Through considerations on the energetics of the droplet (originally in the $m(h)$ system \cite{Binder1980Critical}) it results that, given a finite system size $N$ there is an energy $\e^*(N)$  above which there is no droplet but a supersaturated (homogeneous) phase called {\it evaporated}. Below $\e^*(N)$, the {\it condensed phase} is characterized by the coexistence between the droplet and the supersaturated bulk. The EC transition is such that $\e^*$ and $\beta^*=\beta(\e^*,N)$ converge to the phase transition  values $\ed$, $\betad$ for large sizes, recovering the Maxwell construction (see fig. \ref{fig:quantities}). The convergence is characterized by exponents that we will call $a$ and $b$: 

\barray
\delta\e^*\equiv \ed-\e^*\sim N^{-a} \qquad \delta\beta^*\equiv \beta^*-\betad \sim N^{-b}
\label{eq:abexponents}
.
\earray
When expressed in terms of the size-rescaled control parameter, the EC transition can be considered as a first-order phase transition under some respects, such that the discontinuity is presented by the intensive parameter \cite{Binder2003Theory}. According to the Ising EC theory for the $\beta \leftrightarrow h$, $\e \leftrightarrow m$ system, it is $a=b=1/(d+1)$, in $d$ dimensions. Moreover, the emerging  minority phase droplet at the EC transition is known to exhibit a universal fraction of the system volume, $f=2/(d+1)$ \cite{Biskup2002}.  

After the seminal work  \cite{Binder1980Critical}, there has been a quite intense research activity concerned with the topic. The EC transition and its finite-size rounding have been rigorously  characterized \cite{Biskup2002,Neuhaus2003,Binder2003Theory}. Numerical tests of the EC theory have been performed for different systems in two and three dimensions: the lattice gas at fixed temperature \cite{Neuhaus2003,Nussbaumer2006,Nussbaumer2008} and density \cite{Martinos2007,Zierenberg2015}; the Ising model in the micromagnetic ensemble \cite{Pleimling2001,Kastner2009}; the Lennard-Jones gas at fixed temperature \cite{MacDowell2004,MacDowell2006,Schrader2009} and density \cite{Zierenberg2015}. The general outcome is that the theory provides an accurate description of the EC transition and its finite-size rounding in the Ising-liquid/vapor paradigm.\footnote{Although not analyzed in therms of the Ising EC theory, the EC transition has also been observed in the crystallization of hard spheres \cite{Fernandez2012}.}  

There has also been a variety of studies concerning the Potts model (PM) first-order transition in the microcanonical ensemble, although the situation in this case is much less clear.  Differently from the Ising case, the transition is order/disorder, temperature (not field)-driven. In the nineties, the transition temperatures and interface tension \cite{Gross1996Microcanonical,Janke1998Canonical} were analyzed in 2D with the Metropolis algorithm. Using multicanonical simulations, these quantities were also studied in 3D, up to linear sizes $L=30$ and for $q$ up to $10$ \cite{Bazavov2008} and, for the first time, they estimated the ``spinodal'' interval $\delta\beta^*$, claimed to shrink as $\sim N^{-1/4}$, compatible with $b=1/(d+1)$. In 2007, an efficient cluster algorithm was developed and tested in 2D for $q=10$ up to $L=1024$ \cite{Martin-Mayor2007} (we analyze this data in the present work). Ref. \cite{Troester2012} focuses on the calculation of the interface free energy in the EC transition, and on the related  exceptionally large finite-size effects. Finally, in refs. \cite{Nogawa2011Evaporation,Nogawa2011Static}, the Wang-Landau algorithm was applied in 2D with $q=8$ and $20$, up to $L=1024$ and $512$ respectively. By scaling of the quantity $\delta\beta^*$ versus $\delta\epsilon^*$, they concluded  the validity of the Ising EC exponents $a=b=1/3$ \cite{Nogawa2011Evaporation}. Moreover, the fraction of system volume occupied by the droplet at the transition is claimed to be $f=2/3$, again in agreement with the Ising theory. 

We will show, however, that both the outcome of novel simulations for a larger value of $q$, and the data of ref. \cite{Martin-Mayor2007}, seem to be incompatible with the exponent $b=1/3$. Motivated by such a  controversy, we discuss a possible  alternative to the Ising EC exponents for the 2D PM. We first expose simple arguments to describe the EC phenomenology in terms of Potts quantities and of an exponent $\s$ relating the droplet volume and its interface area. We claim how, in 2D, $\s$ may not assume its geometrical value, $1/2$, but rather $2/3$. Finally we analyze the numerical data and show how, up to the simulated sizes, it is compatible with $\s=2/3$ and incompatible with $\s=1/2$ and hence with the Ising EC scaling. The relationship of the EC phenomena with metastability in the canonical ensemble will be finally discussed. 

\section{Model and notations} 
\begin{figure}
\onefigure[width=.9\columnwidth]{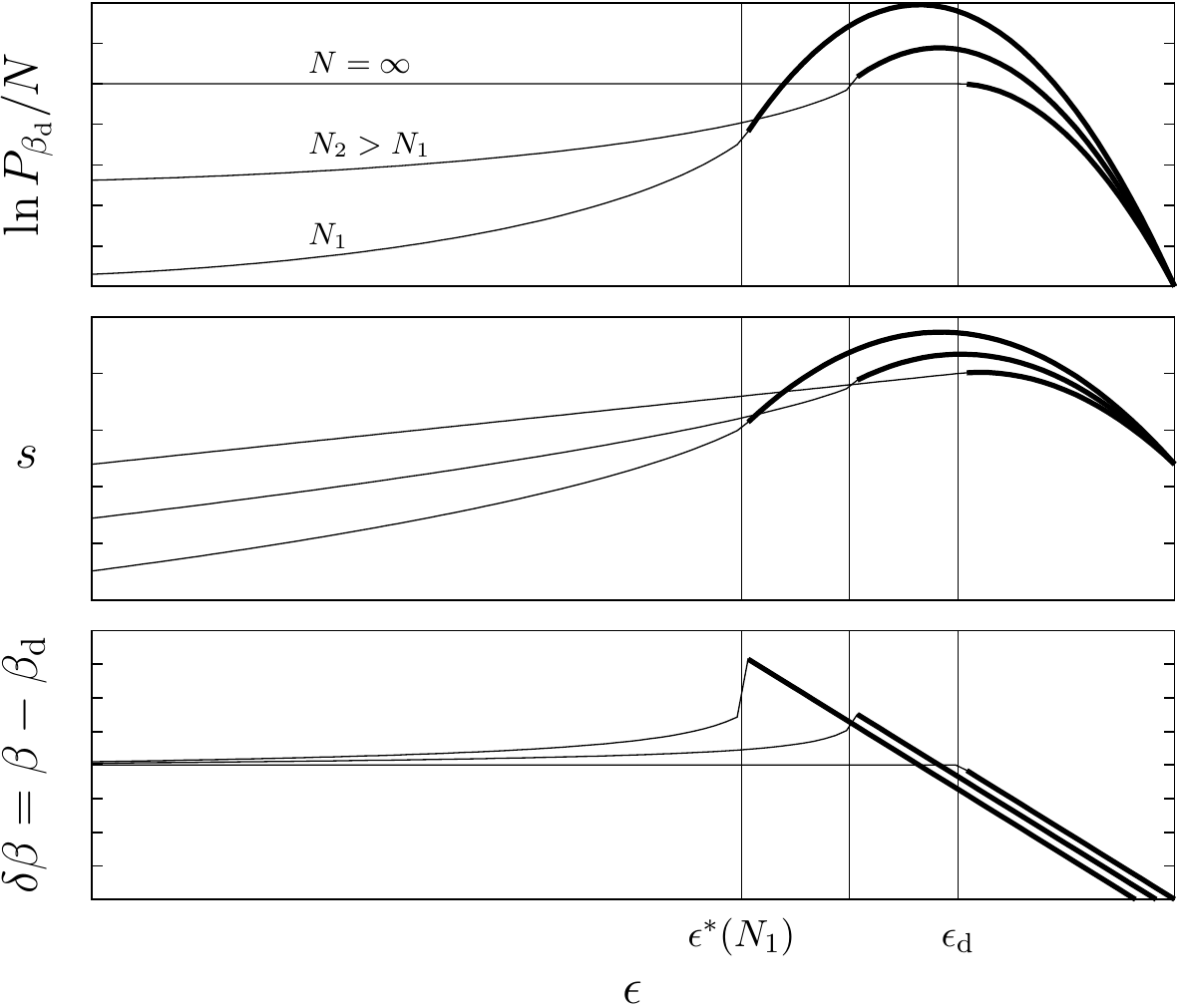}
\caption{Finite-size qualitative behavior of the equilibrium thermodynamic quantities $\ln P/N$ and $s$ (up to a $N$-dependent constant) and $\beta$, versus the intensive energy for a first-order transition in the microcanonical ensemble. The wide and thin curves refer to the evaporated and condensed phases, respectively.}
\label{fig:quantities}
\end{figure}

We consider the $q$-color Potts model in a $d$-dimensional lattice with $N$ sites. It is defined by the Hamiltonian ${\cal H}=\frac{1}{2}\sum_{(i,j)} (1-\delta_{\tau_i,\tau_j})$, where the sum is over the bonds of the lattice, and the degree of freedom $\tau_i$ takes one out of $q$ values. The intensive entropy $s$, inverse temperature, $\beta$, and the (disordered) finite-size energy probability density (EPD) at the transition, $P_\betad$, as functions of $\e,N$, are related in the following way: $ \ln P_\betad/N = -\betad \e + s$, $\beta = \partial_\e s$. For a sufficiently high value of $q>q_{\rm c}(d)$, the transition is first-order \cite{Wu1982Potts}; the quantities $P$, $s$, $\beta$  behave in the coexistence interval qualitatively as sketched in fig. \ref{fig:quantities}.  We will also considered the order and disorder entropies at the transition, $s_{\rm o,d}=s(\e_{\rm o,d},N)$. The quantities $\betad$, $\cd$, $\eo$, $\ed$, $\so$, $\sd$ are $q$, $d$, and lattice-dependent.  \\

\section{Evaporation-condensation in the Potts Model} Our arguments are very similar to that describing the EC-transition in the Ising/Lattice gas model, in the spirit of that of ref. \cite{Nussbaumer2006}. We will focus on the disordered phase, i.e., with $\e$ near the disordered energy $\ed$. 
 
Our first working hypothesis is that the energy difference $\ed-\e$ in the coexistence region can be split into two main contributions: one coming from a {\it single}, connected {\it droplet} of the ordered phase with energy $\eo$ and extensive mass $\lambda N$, $\lambda\in[0:1]$; the second one comes from a disordered, supersaturated fraction of the system, the {\it bulk}, which surrounds the droplet. The equivalent assumption in the Ising case has been  rigorously  proven \cite{Biskup2002}, and for the Potts case it appears to be quite reasonable, at least in two dimensions, in light of the snapshots of the condensed phase configurations shown in the refs. \cite{Martin-Mayor2007,Nogawa2011Evaporation}. The energy difference in the bulk is obtained by an increasing of the correlation length with respect to that at $\ed$.  Our second crucial hypothesis    is that the average interface energy between the droplet and the disordered bulk, proportional to the droplet interface perimeter, is in its turn proportional to a power of the droplet volume: $N \e_{\rm int}=w (\lambda N)^\sigma$, $w$ being the  proportionality constant, and $\s$ being a $d$-dependent exponent, of possible non-geometrical nature \cite{Binder1987Theory,Meunier2000Condensation}:\footnote{
The ambiguity in the droplet definition \cite{Binder1987Theory} reflects in its interface perimeter and area. In the case of geometrical droplets, the interface energy is precisely equal to the interface perimeter of the droplet. We will assume that for different cluster definitions (as the clusters in the Fortuin-Kasteleyn representation) the energy contribution is proportional to the droplet interface, the proportionality constant being $q$ and (slowly) $\e$-dependent, such that it can be absorbed into $w$, not affecting the present discussion. In any case, we expect the droplet definition being less and less influent on the interface energy, the larger the value of $q$. 
}

\beq
\frac{d-1}{d} < \sigma < 1
\eeq 

In this circumstance, the conservation of energy is expressed as 

\beq
\e=\lambda \eo+(1-\lambda)\eb+\e_{\rm int}
\label{eq:nrgconservation}
,
\eeq
where $\eb$ is the energy of the disordered bulk, depending on the droplet volume fraction $\lambda$, with $\e\le\eb\le \ed$.  Due to the presence of the droplet, the bulk energy $\eb$ may be larger than the global energy $\e$, such an energy excess decreases with decreasing size, when the relative contribution of the interface perimeter increases. 

The total entropy   presents contributions from the droplet, bulk and droplet surface, $S_\lambda(\e)=S_{\rm dr}+S_{\rm bu}+S_{\rm sur}$. The droplet entropy is simply $S_{\rm dr} = \lambda N s_o$. The surface entropy is proportional to the droplet perimeter (the {\it contour entropy} $s_{\rm c}$ being the proportionality constant), with a logarithmic correction that we neglect\footnote{The number of self-avoiding polygons with perimeter $p$  is expected to behave as   $a_p\sim c_1^p p^{c_2}$. In any case, neglecting the surface entropy do not alter our conclusions, as we will see.}\cite{Jensen2000Size,Fisher1959Excluded,Jensen2000Statistics}, hence $S_{\rm sur} = s_{\rm c} (\lambda N)^\sigma$. The bulk entropy takes the form $S_{\rm bu} = N\left\{ \sd (1-\lambda) + \betad \zeta-  \betad^2/(2 \cd) \zeta^2 \right\} $, with $\zeta=(1-\lambda) (\eb-\ed)$. It is obtained by expanding up to second order in $\ed-\e$ around the disordered state at $\sd$, where $\cd$ is the specific heat at the transition $\cd=-\betad^2/\partial^2_{\e\e}s(\ed)$. Finally, using $\sd-\so=\Delta \betad$, where $\Delta=\ed-\eo$, and eq. (\ref{eq:nrgconservation}), the total $\lambda$-dependent intensive entropy reads:

\barray
s_\lambda(\e)&=&\sd -\betad \e +C \lambda^\sigma N^{\sigma-1} - \nonumber \\
& - & \frac{1}{2} \frac{\betad^2}{\cd}\Delta^2\left[ \lambda -\frac{\ed-\e}{\Delta} - \frac{w}{\Delta} \lambda^\sigma N^{\sigma-1} \right]^2 
\label{eq:slambda} \\
C&\equiv&\sc-\betad w\nonumber 
\earray

We will call  $\lambda_{\rm m}(\e,N)$ the value maximising $s_\lambda$, such that $s_{\lambda_{\rm m}}$ and $P_{\lambda_{\rm m}}$ become the  finite-size thermodynamic functions, $s$ and $P$. It is expected to be a solution of $\partial_\lambda s_\lambda(\e) =0$ in the condensed phase for $\e$ below a threshold $\e^*(N)$, and $\lambda_{\rm m}=0$ in the evaporated phase for $\e>\e^*(N)$ \cite{Nussbaumer2006}.

{\it Thermodynamic limit.} Maximizing $s_\lambda$ in eq. (\ref{eq:slambda}) with $N=\infty$, one gets $s(\e)=\sd-\betad (\ed-\e)$, i.e., the Maxwell construction, required by van Hove's theorem. The maximizing $\lambda$  corresponds to the {\it lever rule} expected in the thermodynamic limit: $\lambda_{\rm m}=(\ed-\e)/\Delta$.  

{\it Finite sizes.} When the system size $N$ is finite, the entropy gets maximized creating a droplet of size smaller than $(\ed-\e)/\Delta$ due to the $N^{\s-1}$ term in the square bracket in (\ref{eq:slambda}), and to a further penalization for large $\lambda$'s in the $\lambda^\sigma$ term (since $C<0$, as we will see). Looking for the solution of $\partial_\lambda s=0$ one obtains, to second order in $N^{\s-1}$, the equation:

\barray
\lambda -x +n \lambda^{\sigma-1} \left[ \sigma \gamma x -D \right] - \lambda^\sigma \gamma n (1+\sigma) =0
\label{minimumlambdacondition}
\earray 
where we have defined shortcuts for our variables, $\gamma=w/\Delta$, and:

\barray
x=\frac{\ed-\e}{\Delta},\qquad  n=N^{\s-1},\qquad  D=|C|\s \cd/(\betad^2\Delta^2) 
.
\earray

At fixed $x$, there is a value of $n$ (a sufficiently small value of $N$) above which there is no real solution; in other words, for $x<x^*(n)$, $\lambda_m=0$  (the evaporated phase). Conversely, for $x>x^*(n)$ (the condensed phase), $\lambda_{\rm m}\sim x$ is a solution of (\ref{minimumlambdacondition}). From now on, we will refer to $\lambda_{\rm m}$ and to $x^*$ and $\lambda^*$ as functions of the reduced energy and size variables, $(x,n)$.

In the evaporated phase with $\lambda=0$, the $\lambda$-dependent EPD is entirely composed by Gaussian thermal fluctuations $\ln P_{\betad}/N \sim - \betad^2\Delta^2 x^2/(2 \cd)$. In the condensed phase for $x>x^*$ one has $\lambda_{\rm m}\sim r_x x$, being $r_x<1$ a function of $x$ approaching one for large $x$ or low $n$. The EPD presents in this regime an stretched exponential extra term which dominates for large $N$:

\begin{equation}
\frac{\ln P_{\betad}}{N} \sim \left\{ 
\begin{array}{lll}
  - \betad^2\Delta^2 x^2/(2 \cd) & x<x^*(n) & \mathrm{(ev.)}\\
  - |C| (r_xx)^\s N^{\s-1} & x > x^*(n)   & \mathrm{(co.)}
\end{array} \right.
\label{eq:Plambdaregimes}
\end{equation}

An estimation of the $N$-dependence of $x^*$ is obtained equating $\ln P$ in both regimes and, neglecting terms of $\order[N^{2\s-2}]$, one gets 
%
${x^*}^{\s-2}\sim N^{1-\s}$ (the proportionality constant includes $-C^{-1}$, so that  it must be $C<0$ for a solution to exist), which gives the exponent $a$ in eq. (\ref{eq:abexponents}). $b$ is obtained inserting $x^*$ in the expression of $\beta_\lambda$ (derivating $\ln P(\e)_{\betad,\lambda_{\rm m}}$ with respect to $\e$). In substance, 

\beq
a=b=\frac{1-\s}{2-\s}
\label{eq:absigma}
,
\eeq
or $a=b=1/3$ for $\s=1/2$ and $a=b=1/4$ for $\s=2/3$, which is our main prediction. Taking the geometrical value of the sigma exponent, $\s=(d-1)/d$, one recovers the  standard Ising EC exponents $a=b=1/(d+1)$. There is, indeed, essentially the same nucleation-like competition in both cases, between (higher) bulk fluctuations, and (lower) bulk fluctuations plus droplet fluctuations (see for example eq. (7) of \cite{Nussbaumer2006})\footnote{The difference is that, in the Ising case, the surface and bulk free energy terms are independent, while in the Potts case the surface enters as an entropic term by its own, and indirectly in the $\eb$-dependence of the bulk entropy through eq. (\ref{eq:nrgconservation}).}.

\section{Analysis of the solution in two dimensions}
\subsection{The value of the exponent $\s$} In the rest of the article we will focus in the 2D case. Ref. \cite{Meunier2000Condensation} presents an effective droplet (Fisher) series for the 2D PM disordered phase. The results of this  work suggest that the exponent $\s$ could be different from its geometrical value, at least in two dimensions. In the droplet series, the free energy of clusters of area $\ell$ is assumed to be proportional to $\omega \ell^{\sigma_{\rm d}}$, with $\sigma_{\rm d}=2/3$, and $\omega$ being an effective surface tension. The value of $\sigma_{\rm d}$ is obtained requiring the matching with exact results on the free energy cumulants for $q_{\rm c}=4$, or fitting it from the numerical free energy cumulants for hight $q$. The resulting Fisher series accurately describes the energy histograms at $\betad$ (see \cite{Bhattacharya1993Free,Bhattacharya1995Large,Janke1997Monte}). The result $\sigma_{\rm d}=2/3$ is compatible with setting both the energy and the entropy of droplets proportional to the 2/3-th power of their area at fixed energy, as we do. The analogy between the analysis in \cite{Meunier2000Condensation} and the EC case is, however, not straightforward since the droplet theory is concerned with droplets of the metastable state, whose area is of order $1$, while the ones involved in the EC transition are expected to be of order $N$. 

Droplets in the Ising/Lattice gas model at low temperatures are compact since the free energy is minimized by minimizing the perimeter of the droplet/bulk interface. In the Potts model, condensed droplets of the ordered phase at fixed energy, i.e., at fixed perimeter interface, are those maximizing the entropy. It is possible, although we are not aware of such a result, that the most probable two-dimensional lattice polygon with fixed perimeter $p$ exhibits an area proportional to  $p^{3/2}$. It is known that the {\it average} area of self-avoiding lattice polygons and loops with perimeter $p$ is $A\sim p^{3/2}$ \cite{Entig1990,Cardy1994Mean}.\footnote{We note, however, that the average perimeter of self-avoiding polygons with area $A$ is proportional to $A$.}

\subsection{The EC transition for $\sigma=1/2$ and $\s=2/3$}

For $\sigma=1/2$, eq. (\ref{minimumlambdacondition}) becomes a third-order equation in terms of $\mu = \lambda^{1/2}$, which can be solved analytically (see the details in \cite{BerganzaForthcoming}). There is no real solution for $x<x^*(n)=3 (D n/2)^{2/3}+\order[n]$, and $\lambda^*(n)=(D n/2)^{2/3} +  \order[n^{4/3}]$, hence $a=b=1/3$, as we anticipated. In the whole evaporated region for small enough $n$,  the scaling  $\tilde \lambda(\tilde x) = \lim_{n\to 0} \lambda_{\rm m}(x,n)/x^*(n)$ holds, in terms of the scaling variable $\tilde x=x/x^*(n)$. Moreover, the condensed phase fraction $\lambda^*(n)/x^*(n)$ converges as $n\to 0$ to $f_{\s=1/2}=\tilde \lambda(1)=1/3$ .   

\begin{figure}
\onefigure[width=.95\columnwidth]{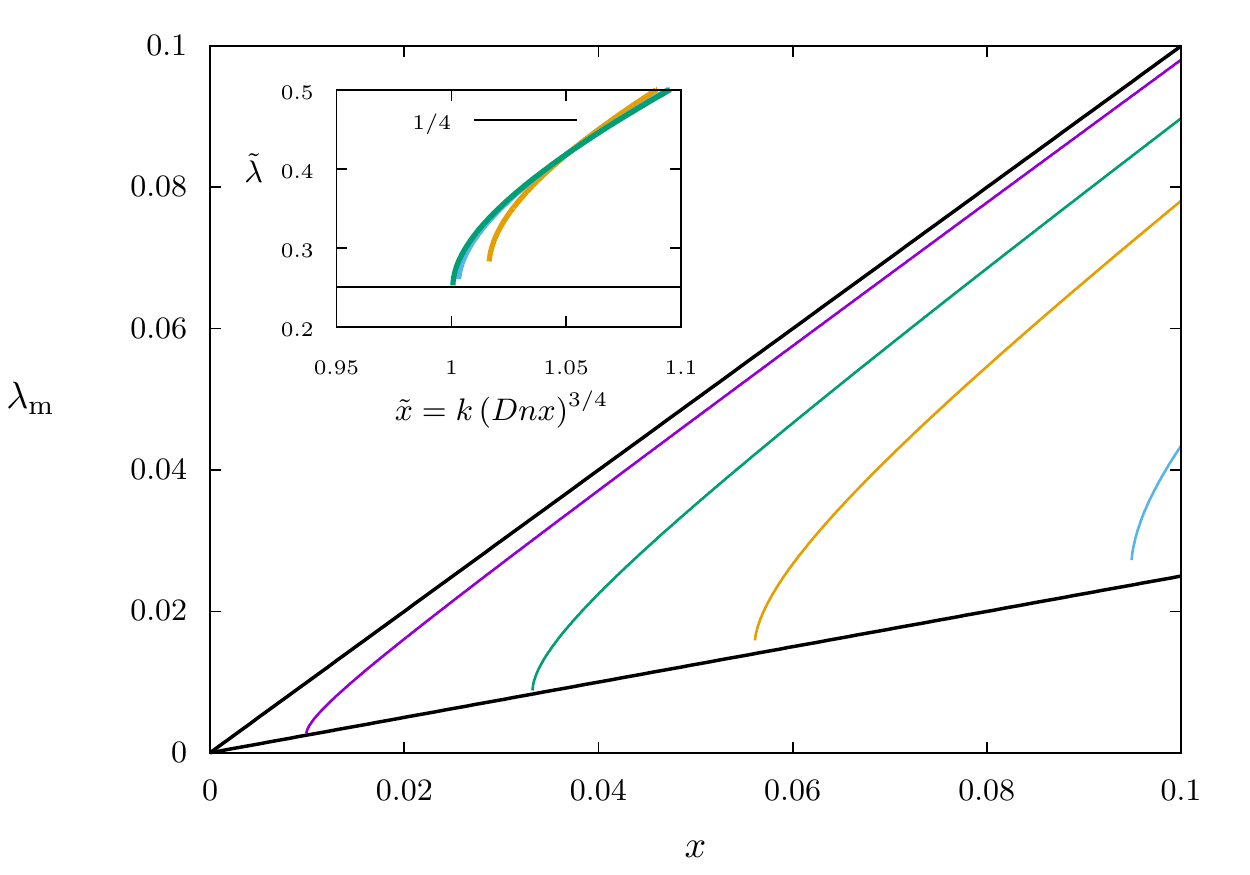}
\caption{The real solution to eq. (\ref{minimumlambdacondition}), for $\s=2/3$, $D=1$ and $n=10^{-2}$, $2\,. 10^{-2}$, $5\, . 10^{-3}$, $10^{-3}$. The thick black lines signal $x$ and $x/4$. Inset: $\tilde\lambda$ versus $\tilde x$ (see main text) for $D=1$, $n=2\, . 10^{-2}$, $2\,. 10^{-3}$; $D=1/2$, $2\,. 10^{-4}$. }
\label{fig:lambda4th}
\end{figure}

In the $\s=2/3$ case, the solution $\lambda_{\rm m}$ is qualitatively identical to that of the $\s=1/2$ case except by the values $a=b=1/4$, as anticipated, and by $f_{2/3}=1/4$. Eq. (\ref{minimumlambdacondition}) takes the form of a fourth order equation in $\mu=\lambda^{1/3}$. Numerically finding the solution of ${\rm Im}[\lambda(n)]=0$, one finds $x^*(n) \sim k\, (n D)^{3/4}+\order[n]$, with $k\simeq 1.7547...$. The discontinuity of the droplet volume fraction turns out to be $f_{2/3}=1/4$ (see the scaling function $\tilde \lambda(\tilde x)$ in fig. \ref{fig:lambda4th}). 

We note that both $f_{1/2}$ and $f_{2/3}$ are different from the value of the equivalent quantity in the Ising case, $2/(d+1)$.

\section{Comparison with numerical data in the microcanonical ensemble}
\label{data}

\begin{figure}
\onefigure[width=.95\columnwidth]{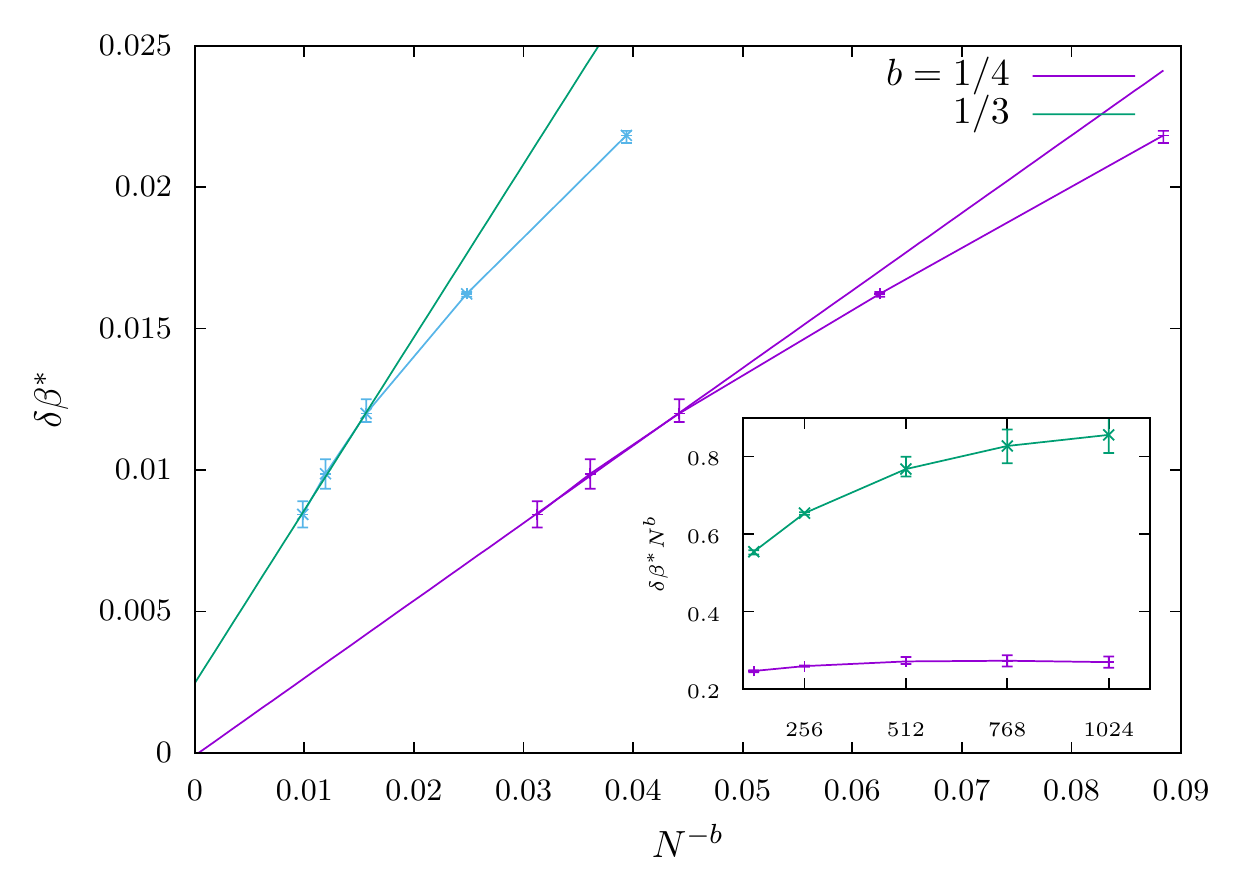}
\caption{Inverse temperature shift of the EC transition $\delta\beta^*$ as a function of $N^{-b}$ for $q=24$, $L=1024$. The intercept term of the linear fit is $y=0.000(4)$ for $b=1/4$, and $y=0.0026(4)$ for $b=1/3$ (the fit interval contains the three largest sizes). Inset: $\delta\beta^*$ multiplied by the $a$ and $b$-th powers of the system size $N$, versus $N^{1/2}$. For $b=1/4$ the three largest sizes are aligned within their errors. }
\label{fig:dbstarq24}
\end{figure}

We have simulated the 2D PM in the microcanonical ensemble with the algorithm presented in ref. \cite{Martin-Mayor2007} (a Monte Carlo cluster algorithm based on the Fortuin-Kasteleyn representation, with Metropolis acceptance probability), for systems with $q=12$, $24$ and up to $L=1024$. The details of our numerical methods will be presented in \cite{BerganzaForthcoming}. We also analyze the data from ref. \cite{Martin-Mayor2007} for $q=10$ up to $L=1024$. For $q=10,12$ we have detected the position of the EC transition point $(\e^*(N),\beta^*(N))$ as the maximum of a c-spline fitting the data. For $q=24$, for which the EC transition becomes sharp, with the prominence of strong  metastable phenomena, we have estimated the lower (upper) bound for $\e^*$ as the highest (lowest) $\e$ at which one observes the condensed (evaporated) phase in a simulated annealing decreasing (increasing) $\e$ \cite{BerganzaForthcoming}, and $\beta^*(N)$ as the $\beta(\e\searrow\e^*(N),N)$ value in the evaporated phase. Plotting the quantity $\delta\beta^*$ versus $N^{-b}$, we observe (see fig. \ref{fig:dbstarq24}) that, up to the simulated sizes, the data are, in principle, compatible with the exponent $b=1/4$ (the linear fit predicts a vanishing $\delta\beta^*(\infty)$, as required by the theory), while they exclude the value $b=1/3$. Moreover, only for $b=1/4$ the quantity $N^b \delta\beta^*$ stays constant for the three largest sizes. For $q=10$, $12$ up to $L=1024$  the situation is qualitatively similar. However, longer simulations are needed for a conclusive confirmation \cite{BerganzaForthcoming}.

 The quantity $\delta\e^* N^a$ reveals more ambiguous results, yet a slight preference for $a=1/4$ is observed \cite{BerganzaForthcoming}. We conclude that the data exhibit a scaling not compatible with the exponent $b=1/3$, although more precise measurements are needed to safely determine the exponents $a$, $b$. 

\begin{figure}
\onefigure[width=.95\columnwidth]{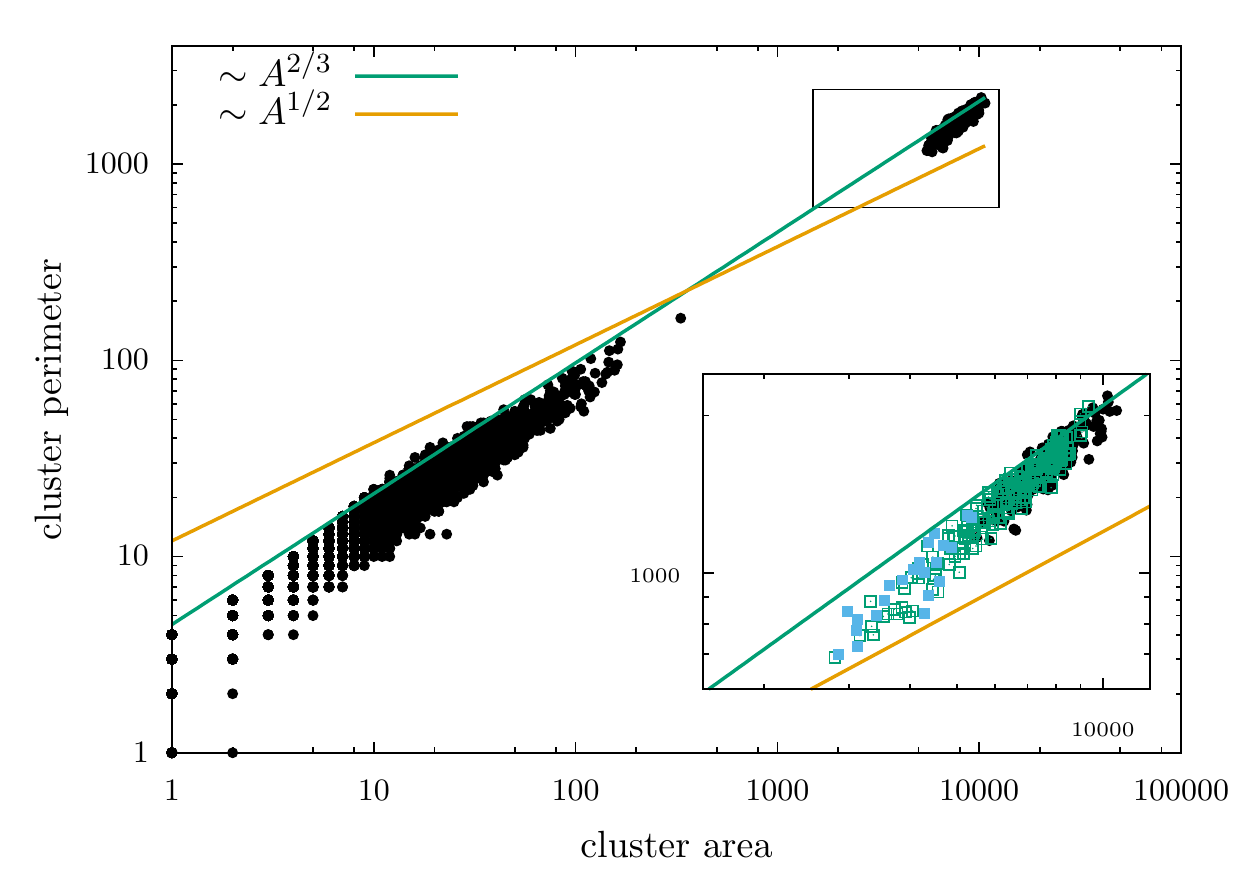}
	\caption{Cluster perimeter versus cluster area ($A$) for $q=24$, $L=1024$, $\e-\ed=-.0158$ (in the condensed phase), along with the functions $4.5\ A^{2/3}$ and $12\ A^{1/2}$. Inset: the macroscopic clusters (in the region signaled by a rectangle in the main figure) are zoomed, and the energies $\e=\ed-.0150$, $\e=\ed-.0142$ added.}
\label{fig:clusters}
\end{figure}

We have also analyzed the statistics of Fortuin-Kasteleyn  cluster areas and perimeters in the condensed phase, where the last quantity is defined as the Potts interface energy of each Wolff cluster \cite{Wolff1989} (both quantities are naturally accessed in the cluster algorithm). The area-perimeter dependence (shown in fig. \ref{fig:clusters} for $q=24$, $L=1024$) provides a more direct indication that the value $\s=2/3$ is more suited for an effective description.

\subsection{Evaporated versus metastable}

The equilibrium curve $\beta(\e)$ in the evaporated phase (illustrated in fig. \ref{fig:quantities}) resembles the metastable continuation in the canonical ensemble. Nevertheless, they are, in general, different. The EC transition converges to the thermodynamical transition in the large-system size limit, while the spinodal point, or the metastable limit, does not in general. In fact, in the Ising model $h$-driven transition, the metastable interval and the lifetime of the metastable phase become independent on size for large sizes \cite{Rikvold1994Metastable,Rikvold1995Recent}.

An opposite example is the 2D Potts model for which, more trivially, metastable states at $\beta>\betad$ can be obtained from a reweighting of the disordered state at $\betad$ \cite{Berganza2014}. The thermodynamics of the metastable state (for $\betad \le \beta \le \beta^*$) is, hence, determined by the finite-size potential, $s(\e,N)$, which is the one fully characterizing also the evaporated phase. We consequently argue that, in this case, the thermodynamic quantities in the evaporated  (microcanonical) phase and in the metastable (canonical) state coincide. Based on such an evaporated/metastable connection, one can estimate the value of  $(\e^*,\beta^*)$ and related quantities  {\it in the metastable state}, if one has a method to systematically exclude out-of equilibrium phenomena (the dynamics of the nucleating droplets), as in ref.  \cite{Berganza2014}. As a confirmation of this fact, we anticipate that the data for $\delta\beta^*$ computed with the present method coincides with that of ref. \cite{Berganza2014} for $q=12$ \cite{BerganzaForthcoming}. Increasing $\delta\beta$ one can in principle approach as much as one likes the point $\beta^*$ and sample more easily the inflexion point of the EPD. The efficiency of such a method  will be analyzed in a forthcoming communication.


\section{Conclusions and Perspectives}

We have presented a phenomenological theory describing the EC transition of the 2D PM. In the presence of minority phase droplets with fractal interface, the EC phenomena result different from that of the known Ising paradigm. This hypothesis is supported by the $\beta(\e)$ data for the 2D PM up to $q=24$, $L=1024$, that we have estimated with the help of a cluster algorithm, although the numerical evidences are still not conclusive (a more accurate estimation of the quantity $\delta\beta^*(N)$ is needed). 


As further perspectives, we propose a quantitative comparison between data and eq. (\ref{eq:Plambdaregimes}): in 2D the relevant quantities $\betad$, $\Delta$, $\cd$ are known \cite{Wu1982Potts}, and $s_{\rm c}$, $w$ could be estimated from geometrical measures of self-avoiding random walks, from measures of the cluster size distribution, or related with the known interface tension of the 2D PM. A more direct test would be a numerical estimation of $\lambda^*/x^*$, to discriminate between the values $f=1/3$, $1/4$, and $2/3$ (see  \cite{Nogawa2011Evaporation}).  A different challenge for future research could be that of clarifying the nature of the 3D system, for which the conventional EC exponents were proposed \cite{Bazavov2008}. A final controversy regarding metastability: we suggest that the thermodynamic metastable spinodal \cite{Berganza2014Anomalous} can be identified with the EC transition; this would contradict the results based on the pseudo-critical divergence \cite{Loscar2009,Ferrero2012,Ferrero2007Longterm}, which indicate that there is a ``thermodynamic'' endpoint of the metastable phase, not vanishing for large system sizes.

A fundamental question, already put forward in ref. \cite{Nogawa2011Static}, is under what general conditions the metastable behavior of a system and its evaporated phase in the micro-ensemble coincide. In this work we have proposed (as suggested in \cite{Nogawa2011Static}) that both coincide in the case of the Potts model, differently with respect to the field-driven Ising transition.

\acknowledgments
I acknowledge Fabrizio Antenucci, Kurt Binder and Alberto Petri for their comments on the manuscript and bibliographic advices, and Tomoaki Nogawa for sharing with me the $b=1/4$ scalings of his data. Particular thanks to V\'ictor Mart\'in-Mayor for useful conversations on the very genesis of the article and for the raw data of reference \cite{Martin-Mayor2007}.
\bibliography{potts}

\begin{thebibliography}{10}
\expandafter\ifx\csname url\endcsname\relax\def\url#1{\texttt{#1}}\fi

\bibitem{Binder1987Theory}
\Name{Binder K.} \REVIEW{Reports on Progress in Physics}{50}{1987}{783+}.

\bibitem{Gunton1983Phase}
\Name{Gunton J., San~Miguel M., Sahni P.~S., Domb C. \and Lebowitz J.}
  \Book{Phase transitions and critical phenomena} (1983).

\bibitem{Fisher1967Theory}
\Name{Fisher M.~E.} \REVIEW{Physics (N. Y.)}{3}{1967}{255}.

\bibitem{Binder1980Critical}
\Name{Binder K. \and Kalos M.} \REVIEW{Journal of Statistical
  Physics}{22}{1980}{363}.

\bibitem{Biskup2002}
\Name{Biskup M., Chayes L. \and Koteck\'y R.} \REVIEW{EPL (Europhysics
  Letters)}{60}{2002}{21}.

\bibitem{Neuhaus2003}
\Name{Neuhaus T. \and Hager J.} \REVIEW{Journal of Statistical
  Physics}{113}{2003}{47}.

\bibitem{Nussbaumer2006}
\Name{Nussbaumer A., Bittner E., Neuhaus T. \and Janke W.} \REVIEW{EPL
  (Europhysics Letters)}{75}{2006}{716}.

\bibitem{Binder2003Theory}
\Name{Binder K.} \REVIEW{Physica A: Statistical Mechanics and its
  Applications}{319}{2003}{99 }.

\bibitem{Nussbaumer2010}
\Name{Nussbaumer A., Bittner E., Neuhaus T. \and Janke W.} \REVIEW{Physics
  Procedia}{7}{2010}{52 } computer Simulation Studies in Condensed Matter
  Physics XX, CSP-2007Proceedings of the Twentieth WorkshopComputer Simulation
  Studies in Condensed Matter Physics XX, CSP-2007.

\bibitem{Nussbaumer2008}
\Name{Nu{\ss}baumer A., Bittner E. \and Janke W.} \REVIEW{Phys. Rev.
  E}{77}{2008}{041109}.

\bibitem{Martinos2007}
\Name{Martinos S., Malakis A. \and Hadjiagapiou I.} \REVIEW{Physica A:
  Statistical Mechanics and its Applications}{384}{2007}{368 }.

\bibitem{Zierenberg2015}
\Name{Zierenberg J. \and Janke W.} \REVIEW{Phys. Rev. E}{92}{2015}{012134}.

\bibitem{Pleimling2001}
\Name{Pleimling M. \and Hüller A.} \REVIEW{Journal of Statistical
  Physics}{104}{2001}{971}.

\bibitem{Kastner2009}
\Name{Kastner M. \and Pleimling M.} \REVIEW{Phys. Rev.
  Lett.}{102}{2009}{240604}.

\bibitem{MacDowell2004}
\Name{MacDowell L.~G., Virnau P., Müller M. \and Binder K.} \REVIEW{The
  Journal of Chemical Physics}{120}{2004}{5293}.

\bibitem{MacDowell2006}
\Name{MacDowell L.~G., Shen V.~K. \and Errington J.~R.} \REVIEW{The Journal of
  Chemical Physics}{125}{2006}{}.

\bibitem{Schrader2009}
\Name{Schrader M., Virnau P. \and Binder K.} \REVIEW{Phys. Rev.
  E}{79}{2009}{061104}.

\bibitem{Fernandez2012}
\Name{Fern\'andez L.~A., Mart\'{\i}n-Mayor V., Seoane B. \and Verrocchio P.}
  \REVIEW{Phys. Rev. Lett.}{108}{2012}{165701}.

\bibitem{Gross1996Microcanonical}
\Name{Gross D. H.~E., Ecker A. \and Zhang X.~Z.} \REVIEW{Annalen der
  Physik}{508}{1996}{446}.

\bibitem{Janke1998Canonical}
\Name{Janke W.} \REVIEW{Nuclear Physics B - Proceedings
  Supplements}{63}{1998}{631}.

\bibitem{Bazavov2008}
\Name{Bazavov A., Berg B.~A. \and Dubey S.} \REVIEW{Nuclear Physics
  B}{802}{2008}{421 }.

\bibitem{Martin-Mayor2007}
\Name{Martin-Mayor V.} \REVIEW{Phys. Rev. Lett.}{98}{2007}{137207}.

\bibitem{Troester2012}
\Name{Troester A. \and Binder K.} \REVIEW{Journal of Physics: Condensed
  Matter}{24}{2012}{284107}.

\bibitem{Nogawa2011Evaporation}
\Name{Nogawa T., Ito N. \and Watanabe H.} \REVIEW{Phys. Rev.
  E}{84}{2011}{061107}.

\bibitem{Nogawa2011Static}
\Name{Nogawa T., Ito N. \and Watanabe H.} \REVIEW{Physics
  Procedia}{15}{2011}{76 } proceedings of the 24th Workshop on Computer
  Simulation Studies in Condensed Matter Physics (CSP2011).

\bibitem{Wu1982Potts}
\Name{Wu F.~Y.} \REVIEW{Reviews of Modern Physics}{54}{1982}{235}.

\bibitem{Meunier2000Condensation}
\Name{Meunier J.~L. \and Morel A.} \REVIEW{The European Physical Journal B -
  Condensed Matter and Complex Systems}{13}{2000}{341}.

\bibitem{Jensen2000Size}
\Name{Jensen I.} \REVIEW{Journal of Physics A: Mathematical and
  General}{33}{2000}{3533}.

\bibitem{Fisher1959Excluded}
\Name{Fisher M.~E. \and Sykes M.~F.} \REVIEW{Phys. Rev.}{114}{1959}{45}.

\bibitem{Jensen2000Statistics}
\Name{Jensen I. \and Guttmann A.~J.} \REVIEW{Journal of Physics A: Mathematical
  and General}{33}{2000}{L257}.

\bibitem{Bhattacharya1993Free}
\Name{Bhattacharya T., Lacaze R. \and Morel A.} \REVIEW{EPL (Europhysics
  Letters)}{23}{1993}{547+}.

\bibitem{Bhattacharya1995Large}
\Name{Bhattacharya T.} \REVIEW{Nuclear Physics B}{435}{1995}{526}.

\bibitem{Janke1997Monte}
\Name{Janke W. \and Kappler S.} \REVIEW{Journal de Physique I}{7}{1997}{663}.

\bibitem{Entig1990}
\Name{Enting I. \and Guttmann A.} \REVIEW{Journal of Statistical
  Physics}{58}{1990}{475}.

\bibitem{Cardy1994Mean}
\Name{Cardy J.} \REVIEW{Phys. Rev. Lett.}{72}{1994}{1580}.

\bibitem{BerganzaForthcoming}
\Name{Ib{\'a}{\~n}ez-Berganza M.} \REVIEW{In preparation}{}{2016}{}.

\bibitem{Wolff1989}
\Name{Wolff U.} \REVIEW{Phys. Rev. Lett.}{62}{1989}{361}.

\bibitem{Rikvold1994Metastable}
\Name{Rikvold P.~A., Tomita H., Miyashita S. \and Sides S.~W.} \REVIEW{Phys.
  Rev. E}{49}{1994}{5080}.

\bibitem{Rikvold1995Recent}
\Name{Rikvold P.~A. \and Gorman B.~M.} \Book{Recent results on the decay of
  metastable phases} 1995 Ch.~5 pp. 149--191.

\bibitem{Berganza2014}
\Name{Ib\'a\~nez Berganza M., Petri A. \and Coletti P.} \REVIEW{Phys. Rev.
  E}{89}{2014}{052115}.

\bibitem{Berganza2014Anomalous}
\Name{Berganza M.~I., Coletti P. \and Petri A.} \REVIEW{EPL (Europhysics
  Letters)}{106}{2014}{56001}.

\bibitem{Loscar2009}
\Name{Loscar E.~S., Ferrero E.~E., Grigera T.~S. \and Cannas S.~A.} \REVIEW{The
  Journal of Chemical Physics}{131}{2009}{024120}.

\bibitem{Ferrero2012}
\Name{Ferrero E.~E., Francesco J. P.~D., Wolovick N. \and Cannas S.~A.}
  \REVIEW{Computer Physics Communications}{183}{2012}{1578 }.

\bibitem{Ferrero2007Longterm}
\Name{Ferrero E.~E. \and Cannas S.~A.} \REVIEW{Physical Review
  E}{76}{2007}{031108+}.

\end{thebibliography}
\bibliographystyle{eplbib}

\end{document}